\documentclass[aps,twocolumn,prl]{revtex4}
\usepackage{graphicx}

\begin{document}
\title{Metal nanoparticles in strongly confined beams: transmission, 
reflection and absorption}
\author{Nassiredin M.\ Mojarad}
\affiliation{Laboratory of Physical Chemistry, ETH Zurich, 8093 Zurich, 
Switzerland}
\author{Gert Zumofen}
\affiliation{Laboratory of Physical Chemistry, ETH Zurich, 8093 Zurich, 
Switzerland}
\author{Vahid Sandoghdar}
\affiliation{Laboratory of Physical Chemistry, ETH Zurich, 8093 Zurich, 
Switzerland}
\author{Mario Agio}
\email{mario.agio@phys.chem.ethz.ch}
\affiliation{Laboratory of Physical Chemistry, ETH Zurich, 8093 Zurich, 
Switzerland}

\keywords{metal nanoparticles, focused light, surface 
plasmon-polaritons, scattering and absorption}

\begin{abstract}
We investigate the interaction of tightly focused light with the 
surface-plasmon-polariton resonances of metal nanospheres.
In particular, we compute the scattering and absorption 
ratios as well as transmission and reflection coefficients. 
Inspired by our previous work in Ref.~\cite{zumofen08}, we discuss how 
well a metal nanoparticle approximates a point-like dipolar radiator. We 
find that a 100 nm silver nanosphere is very close to such an ideal 
oscillator. Our results have immediate implications for single nanoparticle 
spectroscopy and microscopy as well as plasmonics.
\end{abstract}

\maketitle

\section{Introduction}

The use of tightly focused light to investigate individual nano-objects 
has received a renewed attention from both the 
theoretical~\cite{vanenk04,mojarad08,lerme08a,pinotsi08,zumofen08,stobinska08} 
and the experimental points of 
view~\cite{sonnichsen00,boyer02,lindfors04,arbouet04,berciaud04,ewers07,gerardot07,vamivakas07,wrigge08a,wrigge08b,tey08}.
The driving idea is to improve light-matter interaction without the need for 
an optical cavity and to explore the detection limits of very small 
material bodies in free space. Promising results and applications have 
been reported for instance in ultra-sensitive label-free detection of 
biological samples~\cite{ewers07}, spectroscopy of very small gold 
nanoparticles 
(NPs)~\cite{sonnichsen00,boyer02,lindfors04,arbouet04,berciaud04}, 
and of single 
quantum emitters~\cite{gerardot07,vamivakas07,wrigge08a,wrigge08b,tey08,kukura08}.

We have recently shown that a point-like oscillating dipole strongly 
attenuates a tightly confined beam incident on it~\cite{zumofen08}.
In particular, we have found that a focused plane wave can be reflected 
by up to 85\%. These results are valid for a classical 
radiator as well as for a two-level system under weak excitation.
Experiments to confirm these theoretical findings are difficult to
perform on quantum emitters for several reasons. For example, in the 
solid state they should be cooled down to cryogenic 
temperatures~\cite{gerardot07,vamivakas07,wrigge08a,wrigge08b,tey08}, where
high-NA optics is difficult to use. 
On the other hand, handling single atoms in vacuo is difficult because they 
must be trapped and kept in the focal spot~\cite{tey08}.
Here we show that one could think of replacing the quantum emitter with a 
mesoscopic system that has the properties of an oscillating dipole.

Metal NPs have optical properties analogous to a dipolar 
oscillator. They exhibit distinct 
resonances even if their size is smaller than the incident wavelength
because of the excitation of surface plasmon-polaritons~\cite{bohren83}.
In fact, we have recently performed experiments where 
gold NPs were used to investigate fundamental physical 
problems related to classical and quantum oscillators 
close to neutral bodies or interacting with each 
others~\cite{hakanson08,kalkbrenner05}.

Metal NPs, specifically gold and silver, have been studied under tight 
illumination and their plasmon-polariton spectra have been discussed for 
different focusing angles both in the near- and far-field 
regimes~\cite{mojarad08}.
It was shown that higher-order modes are not excited when the 
incident light is tightly focused. On the other hand, it turns out that the 
scattering efficiency at the dipolar resonance is close to the value for 
plane-wave illumination. This quantity is the ratio between the scattering 
cross section and the cross sectional area of the NP and compares the 
electromagnetic size with the geometrical size of the NP~\cite{bohren83}.
Because in a tight focus the intensity is strongly position and
polarization dependent, it is more instructive to adopt another definition 
for the scattering efficiency, namely the scattered power divided by the 
incident power~\cite{zumofen08,mojarad09}. This quantity, which we call 
{\it scattering ratio}, is relevant because, as we will see, it is 
directly related to the transmission of light through the NP and can be 
compared with the result for a point-like oscillating dipole. 
Furthermore, it provides information on the detection limits of very 
small NPs and on the efficiency of light absorption.
For instance, since the background noise is proportional to the incident 
power, increasing the scattering ratio leads to a better signal. 

In this paper we want to focus our attention on the dipolar resonance
and see how well metal NPs behave as point-like dipolar radiators. Moreover,
by computing transmission and reflection we explore the detection limits
for very small NPs and investigate their potential as light absorbers. 
While these topics belong to the basics of light 
scattering~\cite{bohren83}, the treatment of tightly focused light
and power ratios in place of cross sections addresses
questions of current interest.

\section{Results and discussion}

\subsection{Generalized Mie theory}

The problem we want to tackle is a nano-object illuminated by a plane 
wave polarized along $x$ and focused by an aplanatic 
system~\cite{zumofen08,mojarad08,richards59} as 
shown in the inset of Fig.~\ref{ext-r50}. The focused beam can be expressed 
using the multipole expansion
\begin{equation}
\label{incpw}
\mathbf{E}_\mathrm{inc}=\sum_l A_l \left(\mathbf{N}_{e1l}^{(1)}
-i\mathbf{M}_{o1l}^{(1)}\right),
\end{equation}
where $l$ runs from 1 to infinity, and $\mathbf{N}_{e1l}^{(1)}$,
$\mathbf{M}_{o1l}^{(1)}$ are respectively the transverse magnetic and 
transverse electric multipoles according to the notation of 
Ref. ~\cite{bohren83}. $A_l$ are the expansion 
coefficients~\cite{sheppard97,mojarad08}
\begin{eqnarray}
A_l & = & (-i)^lE_0kf\frac{2l+1}{2l^2(l+1)^2}\times \\ \nonumber
& & \int_0^\alpha a(\theta)
\left[\frac{P_l^1(\cos\theta)}{\sin\theta}+
\frac{\mathrm{d}P_l^1(\cos\theta)}{\mathrm{d}\theta}\right]
\sin\theta\mathrm{d}\theta,
\end{eqnarray}
where $E_0$ is the amplitude of the incident electric field, $f$ is the 
focal length of the lens, $k$ is the wavevector and 
$a(\theta)=\sqrt{\cos\theta}$ is the apodization function for an aplanatic 
system~\cite{richards59}. Note that the focusing angle $\alpha$ is the only 
parameter that dictates the relative weight $\left|A_l/A_1\right|$.

In the next step a sphere is placed at the focus. The 
scattered field $\mathbf{E}_\mathrm{sca}$ can be found by 
matching the boundary conditions at its surface
\begin{equation}
\label{scapw}
\mathbf{E}_\mathrm{sca}=\sum_l A_l \left(-a_l \mathbf{N}_{e1l}^{(3)}
+ib_l \mathbf{M}_{o1l}^{(3)}\right),
\end{equation}
where $a_l$ and $b_l$ are the Mie 
coefficients~\cite{bohren83,mojarad08}. As soon as the incident and 
scattered fields are determined, the scattered power ($P_\mathrm{sca}$) 
and the sum of scattered and absorbed powers ($P_\mathrm{s+a}$) are 
calculated by respectively integrating the Poynting vectors 
of the scattered field and of the interference term
over the entire space. Using the properties of multipoles, 
analytical expressions for $P_\mathrm{sca}$ and $P_\mathrm{s+a}$ are 
found to be~\cite{mojarad08}
\begin{eqnarray}
\label{ws}
P_\mathrm{sca} & = & \frac{\pi}{2Zk^2}\sum_l
\frac{2l^2\left(l+1\right)^2}{2l+1}\left|A_l\right|^2
\left(\left|a_l\right|^2+\left|b_l\right|^2\right),\\
\label{we}
P_\mathrm{s+a} & = & \frac{\pi}{2Zk^2}\sum_l
\frac{2l^2\left(l+1\right)^2}{2l+1}\left|A_l\right|^2
\mathrm{Re}\left\{a_l+b_l\right\},
\end{eqnarray}
where $Z$ is the impedance in the background medium.
Note that Eq.~(\ref{we}) is the textbook formula for the 
extinction~\cite{bohren83}, but $P_\mathrm{s+a}$ does not represent 
the power removed from the incident beam in a transmission experiment, 
because for focused illumination the optical theorem is not 
valid~\cite{lock95b}. We discuss this point in
Sec.~\ref{transmission}, where we consider the problem of transmission and 
reflection. The absorbed power is
defined as usual by $P_\mathrm{abs}=P_\mathrm{s+a}-P_\mathrm{sca}$.

\subsection{Scattering and absorption ratios}

Contrary to a plane wave, a beam carries a finite amount of power. Thus, 
we can quantify the response of our nano-object by considering the 
scattering ($\mathcal{K}_\mathrm{sca}$)
and absorption ($\mathcal{K}_\mathrm{abs}$) ratios that are  
defined by dividing $P_\mathrm{sca}$ and $P_\mathrm{abs}$ respectively 
by the total incident power $P_\mathrm{inc}$~\cite{zumofen08,mojarad09},
\begin{equation}
P_\mathrm{inc}=\frac{f^2}{2Z}
\int_0^{2\pi}\int_{\pi/2}^{\pi}
\left|\mathbf{E}_\mathrm{inc}\cdot\mathbf{n}\right|^2
\sin\theta\,\mathrm{d}\theta\,\mathrm{d}\phi,
\end{equation}
where $\mathbf{E}_\mathrm{inc}$ is taken in the far field at $f$
and $\mathbf{n}$ is the unit vector normal to the 
spherical wave-front. For completeness, we also define the quantity 
$\mathcal{K}_\mathrm{s+a}=P_\mathrm{s+a}/P_\mathrm{inc}$.
We focus our attention on silver and gold NPs because they exhibit a 
strong resonant behavior in the visible spectral range. 
The NPs are in a homogeneous background medium with index of refraction 
$n_\mathrm{b}=1.3$ (immersion in water).

\begin{figure}[!h]
\begin{center}
\includegraphics[width=8cm]{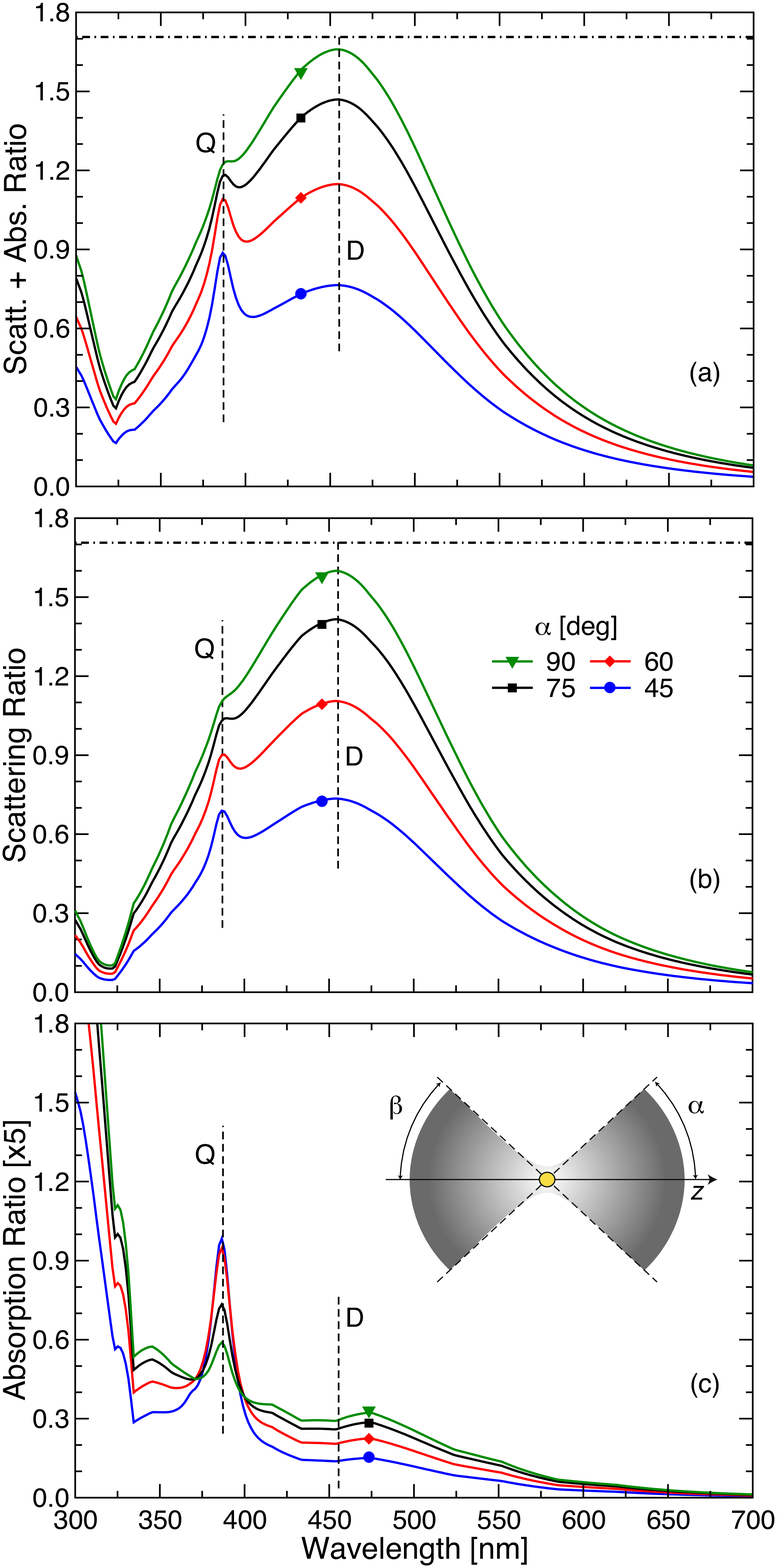}
\caption{\label{ext-r50}(a) Scattering plus absorption 
$\mathcal{K}_\mathrm{s+a}$, (b) scattering $\mathcal{K}_\mathrm{sca}$ and
(c) absorption $\mathcal{K}_\mathrm{abs}$ ratios for a 
100 nm silver NP in water ($n_\mathrm{b}=1.3$) under a focused plane wave
for different focusing angles $\alpha$. The collection angle is 
$\beta=180^\circ$. The dashed-dotted lines in (a) and (b) refer to 
$\mathcal{K}_\mathrm{sca}$ for a resonant point-like oscillating dipole for 
$\alpha=90^\circ$~\cite{zumofen08}. The vertical dashed lines mark the 
dipole (D) and quadrupole (Q) resonances. 
$\mathcal{K}_\mathrm{abs}$ is multiplied by a factor of 5 for clarity.
The inset in (c) shows a NP in a focused beam, indicating the focusing 
angle $\alpha$, the collection angle $\beta$, and the optical axis 
$z$.} \end{center}
\end{figure}

Figure~\ref{ext-r50} displays $\mathcal{K}_\mathrm{s+a}$, 
$\mathcal{K}_\mathrm{sca}$ and $\mathrm{K}_\mathrm{abs}$ for a 100 nm 
silver NP for different focusing angles. Tighter focusing has two major 
effects. First, the efficiency increases and
reaches the maximum value for $\alpha=90^\circ$. Second, a stronger 
focus reduces the coupling to higher order multipoles because they 
have a small contribution in the incident field~\cite{mojarad08}.
As a consequence, $\mathcal{K}_\mathrm{abs}$ decreases at the quadrupole 
(Q) resonance if $\alpha$ increases.
The fact that $\mathcal{K}_\mathrm{s+a}$ and $\mathcal{K}_\mathrm{sca}$ 
can exceed 1 does not violate the conservation 
of energy since $P_\mathrm{sca}$ and $P_\mathrm{s+a}$ contribute to the 
energy balance with opposite signs~\cite{bohren83} and their sum never 
exceeds $P_\mathrm{inc}$~\cite{zumofen08}.

A small metal NP is often treated using a dipolar polarizability
\begin{equation}
\label{polarizability}
\alpha_\mathrm{NP}=\frac{\alpha_\circ}{1-i\frac{k^3}{6\pi}\alpha_\circ},
\end{equation}
where $\alpha_\circ$ is the NP electrostatic 
polarizability~\cite{bohren83,meier83}.
Equation~(\ref{polarizability}) shows that $\alpha_\mathrm{NP}$ approaches 
the value of a classical oscillator on 
resonance~\cite{jackson99,zumofen08} only 
if $\alpha_\circ \gg 6\pi/k^3$. This occurs if
the absorption losses are negligible with respect to radiation, 
irrespective of the magnitude of $\alpha_\circ$, or if $\alpha_\circ$ is 
large because of the NP size. In the first case, the losses can be
reduced by choosing the material composing the NP, or by changing the 
background index and the NP shape to shift the resonance~\cite{bohren83}. In 
the second case, the size of the NP cannot be too large because higher-order
resonances begin to spectrally overlap with the dipolar one and
depolarization effects become stronger~\cite{bohren83}.
Thus, the interesting question that arises is: when do a NP and an 
oscillating dipole behave in a similar way under tight illumination?

The dashed-dotted lines in Fig.~\ref{ext-r50}(a) and \ref{ext-r50}(b) show 
$\mathcal{K}_\mathrm{sca}$ of a resonant point-like dipole for
the focusing angle $\alpha=90^\circ$. 
Also in this case, $\mathcal{K}_\mathrm{sca}$, which we call 
$\mathcal{K}_\mathrm{dp}$ for a point-like dipole, is a function of $\alpha$ 
and reads \begin{equation}
\mathcal{K}_\mathrm{dp}=\frac{128}{75}\frac{1}{\sin^2\alpha}
\left[1-\frac{1}{8}(5+3\cos\alpha)\cos^{3/2}\alpha\right]^2.
\end{equation}
This value is independent of the dipole resonant wavelength
and for $\alpha=90^\circ$ reaches the maximum value of 
$\mathcal{K}_\mathrm{dp}\simeq 1.707$~\cite{zumofen08}.
As depicted in Figure~\ref{ext-r50}(b), a 100 nm silver NP in 
water exhibits a $\mathcal{K}_\mathrm{sca}$ very close to 
$\mathcal{K}_\mathrm{dp}$ on its dipole resonance.
It could also be seen that a 100 nm gold NP does not approach the 
scattering ratio of a perfect radiator as well, because of the 
higher material losses. 

\subsection{Transmission, reflection and absorption}
\label{transmission}
The scattering $\mathcal{K}_\mathrm{sca}$ and absorption 
$\mathcal{K}_\mathrm{abs}$ ratios determine the conversion of the 
incident power into the corresponding 
quantities. Nevertheless, they give no information about the directionality 
of the scattered field. By investigating the transmission and reflection 
of a beam interacting with a nano-object a better understanding of 
the process is obtained.
The simplest and most commonly discussed case is that of a plane
incident wave~\cite{jackson99,bohren83}.
If the power is only collected along the $z$ axis, the transmission 
is expressed as
\begin{equation}
\label{beer}
T=\frac{I}{I_\mathrm{inc}}=1-\frac{\sigma}{\mathcal{A}},
\end{equation}
where $I$ and $I_\mathrm{inc}$ are the transmitted and incident 
intensities, respectively. $\mathcal{A}$ is the beam cross sectional 
area, and $\sigma$ is a cross section. The ratio $\sigma/\mathcal{A}$ in 
Eq.~(\ref{beer}) represents the fraction of the power extinguished from 
the incident beam. In the context of the optical theorem, $\sigma$ 
results from the interference term in Eq.~(\ref{we}), which is thus referred
to as the ``extinction'' cross section. However, if the off-axis
collection is included, a fraction of the scattered power must be
added to Eq.~(\ref{beer})~\cite{bohren83}.

For focused illumination the intensity is not homogeneous and 
the optical theorem no longer holds~\cite{lock95b}. In fact, 
$P_\mathrm{s+a}$ is not all concentrated
along the $z$ axis, but it covers a range of angles from $\beta=0^\circ$ to
$\beta=\alpha$. Thus, $P_\mathrm{s+a}$ does not correspond to the power 
removed from the beam in a transmission experiment. 
For this reason, we do not equate $P_\mathrm{s+a}$ with the extinguished 
power, although this convention can be found in the 
literature~\cite{lock95b,wrigge08b}.
If the collection angle is extended to the forward 
hemisphere, i.e. $\beta=90^\circ$, the transmission ($T$) can be 
nevertheless formulated in terms of the ratio of powers as~\cite{zumofen08}
\begin{equation}
\label{transdip}
T=\frac{P}{P_\mathrm{inc}}=1-\frac{1}{2}\mathcal{K}_\mathrm{dp},
\end{equation}
where $P$ is the transmitted power.

In the case of a dissipating dipole, absorption also plays a major role 
in the amount of light that gets transmitted and the generalized 
formula for tightly focused beams becomes
\begin{equation}
\label{trans}
T=1+\frac{1}{2}\mathcal{K}_\mathrm{sca}-\mathcal{K}_\mathrm{s+a}=
1-\frac{1}{2}\mathcal{K}_\mathrm{sca}-\mathcal{K}_\mathrm{abs}.
\end{equation}
Moreover, scattering and absorption have a different weight, because half 
of the scattered power is added to 
the incident power in the forward direction, while 
$\mathcal{K}_\mathrm{s+a}$ contributes with a negative sign.
Equation~(\ref{trans}) tells us that the correct definition of extinction 
for tight illumination, i.e. the fraction of power removed in the 
forward direction, is
$\mathcal{K}_\mathrm{abs}+\mathcal{K}_\mathrm{sca}/2$.
However, when the power is collected with an angle $\beta<90^\circ$ 
the above expressions have a more complicated form~\cite{zumofen08}.
Furthermore, Eq.~(\ref{trans}) is not valid for a non-dipolar 
scattered, because $P_\mathrm{sca}$ is not equally distributed in the 
forward and backward directions. Consequently, we compute 
the transmission through metal NPs by integrating the Poynting 
vector of the total field in the forward hemisphere ($\beta=90^\circ$).
Reflection ($R$) is given by integrating the Poynting vector of the 
scattered field in the backward hemisphere or alternatively
by $R=1-T-\mathcal{K}_\mathrm{abs}$.

\begin{figure}[!h]
\begin{center}
\includegraphics[width=8cm]{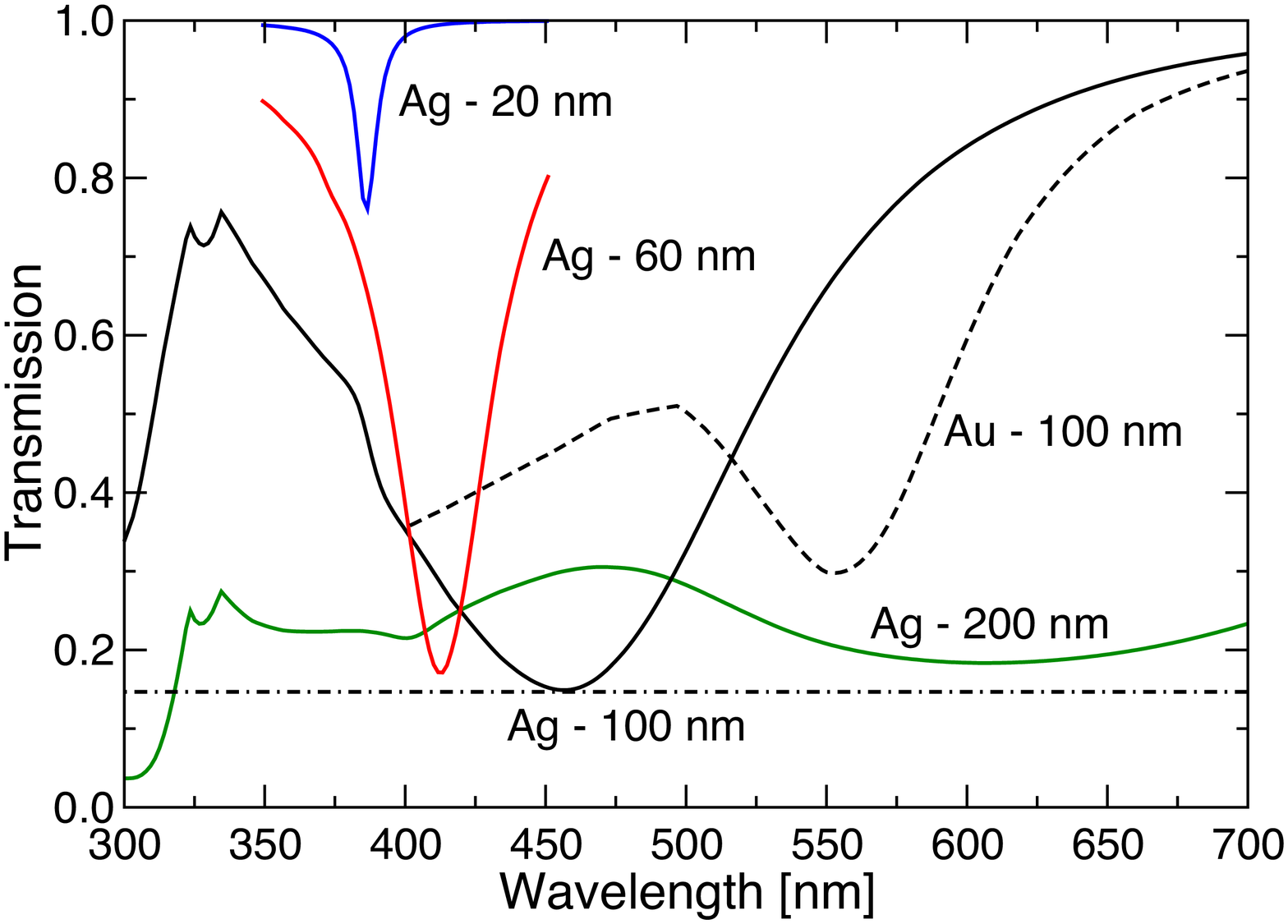}
\caption{\label{ext-trans-alpha90}
Transmission ($\beta=90^\circ$) for silver and gold NPs in water 
($n_\mathrm{b}=1.3$) under a tightly focused plane wave 
($\alpha=90^\circ$).
The dashed-dotted line refers to the transmission for a resonant point-like 
oscillating dipole under the same illumination 
conditions~\cite{zumofen08}.}
\end{center}
\end{figure}

Figure~\ref{ext-trans-alpha90} displays the transmission spectrum for
various metal NPs. We note that a 100 nm silver NP exhibits almost the
same transmission dip of a point-like oscillating dipole when excited at
the dipole resonance. The latter is indicated by a dashed-dotted line and
corresponds to the value given by Eq.~(\ref{transdip}). Indeed,
as shown in Fig.~\ref{ext-r50}, because $\mathcal{K}_\mathrm{sca}\simeq 
\mathcal{K}_\mathrm{dp}$ and $\mathcal{K}_\mathrm{abs}\ll 
\mathcal{K}_\mathrm{sca}$, Eqs.~(\ref{transdip}) and (\ref{trans}) 
shall yield almost the same result.
Even if the dip is very close to that of a point-like oscillating dipole, the 
transmission spectrum of the 100 nm silver NP is not as simple as a
Lorentzian profile~\cite{zumofen08}.
The reason is that the NP supports a quadrupole resonance around 380 nm
and the dipolar polarizability of the NP includes material 
dispersion~\cite{johnson72} and depolarization effects~\cite{meier83}.
In fact, smaller silver NPs yield a transmission spectrum that is closer
to that of a dipolar radiator, but the dip is smaller because 
absorption reduces the strength of the polarizability on resonance.
For the same reason a 100 nm gold NP yields a transmission larger
than a perfect radiator.

On the other hand, larger metal NPs support higher-order resonances that 
spectrally overlap and, for this reason, exhibit a more complex scattering 
pattern. More precisely, because $P_\mathrm{sca}$ is not equally 
distributed in the forward and backward directions, Eq.~(\ref{trans})
is not valid. 
Indeed, Fig.~\ref{ext-trans-alpha90} shows that for a 200 nm silver NP
the transmission spectrum does not resemble a Lorentzian profile and, at the 
dipolar resonance ($\lambda\simeq 600$ nm), the dip is less pronounced than 
for a 100 nm silver NP ($\lambda\simeq 455$ nm).

\begin{figure}[!h]
\begin{center}
\includegraphics[width=8cm]{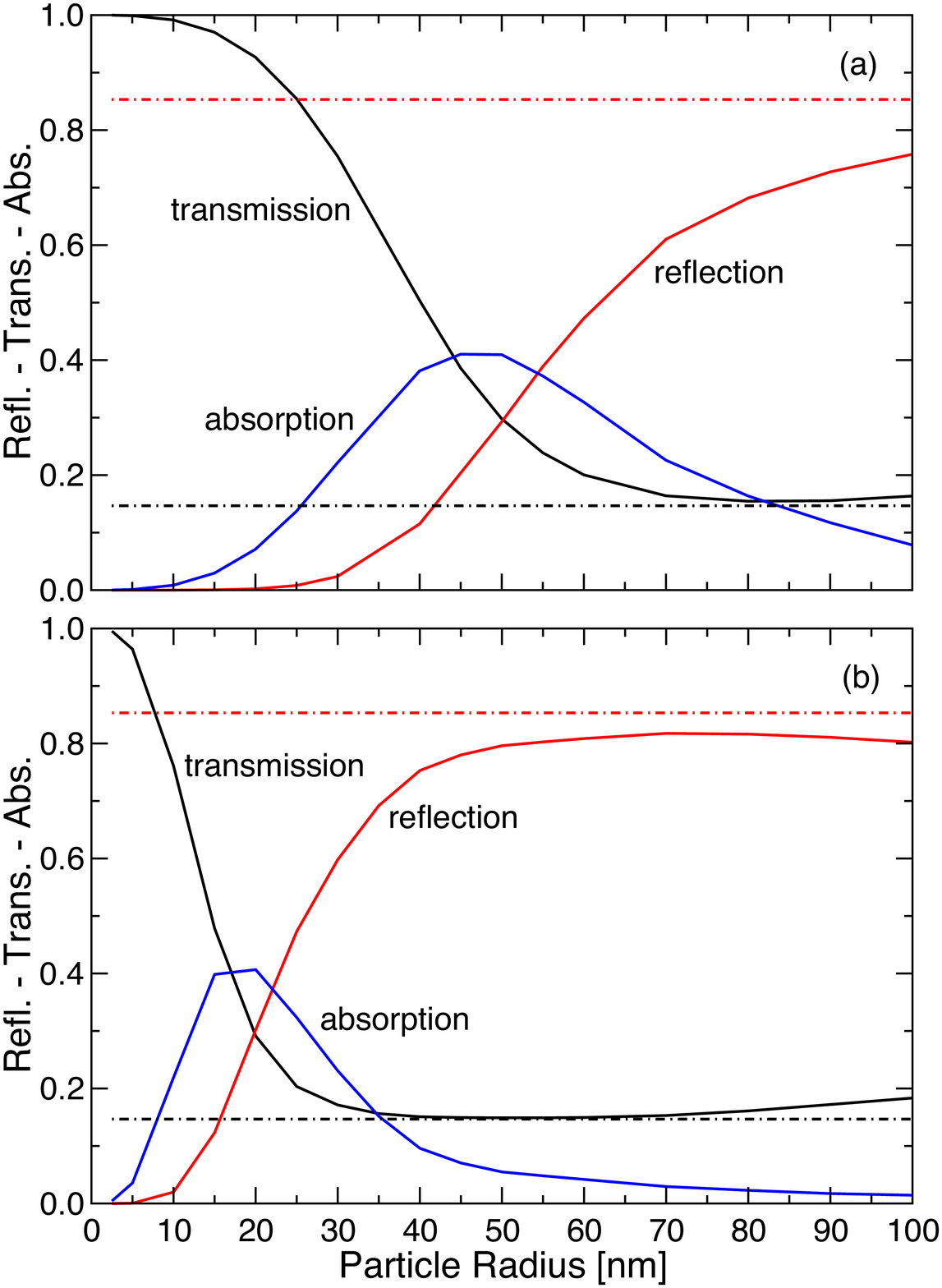}
\caption{\label{trans-alpha90}Transmission, reflection and absorption 
$\mathcal{K}_\mathrm{abs}$ for (a) gold and (b) silver NPs in water 
($n_\mathrm{b}=1.3$) under a tightly focused plane wave 
($\alpha=90^\circ$) as a function of the NP size. 
The dashed-dotted lines refer to transmission and reflection for a 
resonant point-like oscillating dipole under the same illumination 
conditions~\cite{zumofen08}.}
\end{center}
\end{figure}

\subsection{Discussion}

To better understand how close a metal NP to a point-like oscillating 
dipole is, we consider the transmission dip and the reflection peak as a 
function of the NP radius. While for 
silver NPs the transmission dip and the 
reflection peak occur at the same wavelength, 
for gold NP there is a small shift due to the fact that the 
plasmon-polariton resonance crosses the onset of interband 
transitions~\cite{johnson72}, which 
gives a rapid increase of losses and hence reshapes the transmission 
profile. Figure~\ref{trans-alpha90}(a) 
and ~\ref{trans-alpha90}(b) respectively show the results for gold and 
silver NPs. The dashed-dotted lines refer to the maximal reflection and 
transmission for an oscillating dipole~\cite{zumofen08}.
We find that a gold NP approaches a dipolar radiator when its radius is 
larger than 80 nm and that above 90 nm the transmission dip starts
to increase because of the spectral overlap with higher-order modes.
On the contrary, a silver NP reaches the properties of an ideal 
oscillating dipole already for radii as small as 40 nm.

We would like to stress that the transmission dip is not enough to judge 
how close the NP is to a dipolar radiator model. That is because the 
transmission dip contains also the effect of absorption, as shown in 
Eq.~(\ref{trans}), while the reflection only provides information on
scattering. Indeed we find that even when the transmission 
dip is close to the limit of a perfect radiator, 
$\mathcal{K}_\mathrm{abs}$ is 
still significant. It is also interesting to see that for a silver NP 
$\mathcal{K}_\mathrm{abs}$ is maximal when the radius is about 20 
nm, while for gold when the radius is about 45 nm. 
While it is commonly assumed that very small NPs are light absorbers, 
one has to keep in mind that the efficiency of the process goes down
(see Fig.~\ref{trans-alpha90}).
For the purpose of converting a certain amount of input power into heat, 
one sees that the highest efficiency is met when scattering is not small. 
In addition, when the NP radius is smaller than about 10 nm for silver 
NPs and 30 nm for gold NPs, the detection signal is stronger if one 
measures the transmission dip or equally the absorbed 
power~\cite{berciaud04}. Indeed, transmission and absorption give exactly 
the same contrast because the scattered power is negligible and 
Eq.~(\ref{trans}) becomes $T=1-\mathcal{K}_\mathrm{abs}$.
For instance, for very small diameters such as 5 nm, a silver and a gold 
NP would respectively exhibit a contrast of $\simeq$ 0.45\% and $\simeq$ 
0.013\%. 

\begin{figure}[!h]
\begin{center}
\includegraphics[width=8cm]{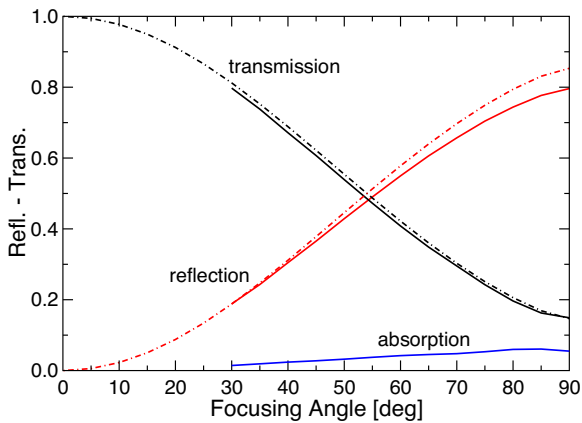}
\caption{\label{trans-r50}Transmission, reflection and absorption 
$\mathcal{K}_\mathrm{abs}$ for a 100 nm silver NP in water 
($n_\mathrm{b}=1.3$) under a tightly 
focused plane wave as a function of the focusing angle $\alpha$. 
The dashed-dotted lines refer to transmission and reflection for a 
resonant point-like oscillating dipole under the same illumination 
conditions~\cite{zumofen08}.}
\end{center}
\end{figure}

To further investigate how a metal NP mimics a point-like 
radiator, we focus our attention on a 100 nm silver NP and examine
the transmission dip and reflection peak as a function of the focusing angle.
The excitation wavelength is fixed to the dipole resonance. 
Figure~\ref{trans-r50} compares these two quantities with the values 
obtained for an ideal oscillating dipole~\cite{zumofen08}. The agreement 
is very good over a broad range of focusing angles. The maximum 
deviation of the reflection curves is at $\alpha=90^\circ$, where it 
reaches about 6\%. This shows that more energy goes into 
absorption when focusing more tightly.

\section{Conclusions}

We studied the scattering properties of spherical silver and gold NPs 
illuminated by a tightly focused plane wave and compared them with the 
response of an ideal oscillating dipole. We have described this 
process by introducing the scattering, and absorption ratios.
These quantities are different from the well 
known scattering and absorption efficiencies for plane wave 
illumination~\cite{bohren83} as well as those defined for a Gaussian 
beam~\cite{lock95a}.
We have also seen that the maximal $\mathcal{K}_\mathrm{abs}$ for silver and 
gold NPs respectively occurs for diameters of 40 nm and 90 nm. 
These results are relevant for applications that exploit the heating 
created by the metal NP, such as in cancer therapy~\cite{huang08}, laser 
ablation and optical data storage~\cite{sugiyama01}.

When the response of the metal NP can be described by a dipolar 
polarizability, Eq.~(\ref{trans}) relates the transmission to
the scattered and absorbed powers.
With that one can have a quantitative estimate of the signal 
from metal NPs~\cite{sonnichsen00,lindfors04,arbouet04,berciaud04} and also
other nano-objects like colloidal quantum dots and 
viruses~\cite{ewers07,kukura08}.

It was shown that a 100 nm silver NP behaves to a very good approximation
like a point-like radiator. This suggests that such NPs could be 
used for fundamental experiments where quantum emitters are replaced by 
classical oscillating dipoles. These include studies on the 
van der Waals-Casimir interactions~\cite{hakanson08}, and on the
extinction properties of emitters~\cite{wrigge08a}. In the latter case,
one could investigate improved focusing systems,
such as directional dipole waves~\cite{zumofen08} and perform simpler test 
experiments at room temperature with metal NPs instead of quantum
emitters at cryogenic temperatures~\cite{wrigge08a}.
Furthermore, one could study 
the coupling of single photons to a plasmon-polariton 
mode~\cite{tame08} with a large efficiency at room temperature.
Finally, our results may find application also in the area of ultrafast 
phenomena and coherent control~\cite{aeschlimann07,stockman07}, where
the interaction of the incoming beam with the system plays a critical role.

\section*{Acknowledgments}
This work was supported by the ETH Zurich research grant TH-49/06-1.

\end{document}